\documentclass{edp-conf}
\usepackage{graphicx}
\begin{document}

\TitreGlobal{SF2A 2005}

\title{Star formation in the merging galaxy cluster Abell 3921$^*$}
\thanks{$^*$This research was supported by
Marie Curie individual fellowship MEIF-CT-2003-900773.}
\author{Ferrari, C.}
\address{Institut f\"ur Astrophysik, Technikerstra{\ss}e 25, 6020
Innsbruck, Austria} 
\author{Maurogordato, S.} \address{Observatoire de la
C\^ote d'Azur, Laboratoire Cassiop\'ee, BP4229, 06304 Nice Cedex 4,
France}
\author{Feretti, L.}\address{INAF, Istituto di Radioastronomia, via Gobetti 101, 40129
Bologna, Italy} 
\author{Hunstead, R.W.}  \address{School of Physics,
University of Sydney, NSW 2006, Australia} 
\author{Benoist, C.$^2$}   
\author{Cappi, A.}
\address{INAF, Osservatorio Astronomico di Bologna, via Ranzani 1,
40127 Bologna, Italy} 
\author{Schindler, S.$^1$}
\author{Slezak, E.$^2$} 
\runningtitle{Star formation in A3921}
\setcounter{page}{237}
\index{Ferrari, C.}
\index{Maurogordato, S.}
\index{Feretti, L.}
\index{Hunstead, R.W.}
\index{Benoist, C.}
\index{Cappi, A.}
\index{Schindler, S.}
\index{Slezak, E.}

\maketitle
\begin{abstract} {Through a combined optical and radio analysis, 
we have investigated the possible connection between the dynamical
state of the merging cluster A3921 and its star formation properties,
reaching the conclusion that the on-going merger is triggering a
SF episode in the collision region.}
\end{abstract}
%
\section{Introduction}

In the past 25 years, optical observations of clusters have revealed
an evolution in galaxy properties with redshift. The fraction of
star-forming/post-star-forming cluster objects significantly increases
with $z$, going from $\sim$1-2\% in the local Universe (Dressler 1987)
to $\geq$30\% at $z{\sim}$0.3-0.5 (Dressler et al. 1999). In 1978
Butcher \& Oemler reported a strong evolution from redder to bluer
colours in cluster galaxies, detecting an excess of blue galaxies at
$z$=0.5 with respect to lower redshift systems. A debate about the
physical origin of this effect began: the observed change of galaxy
colours with $z$ could be simply due to a passive evolution of
infalling field galaxies, or it could be strongly affected by
environmental effects.  In 1983 Dressler \& Gunn pointed out for the
first time that in fact the blue colour of the population detected by
Butcher \& Oemler was the result of SF activity. Since then, many
different studies have tried to understand the origin of the observed
evolution in the SF history of cluster galaxies. Several physical
mechanisms have been proposed that may either trigger or weaken SF
within clusters (e.g. Dressler \& Gunn 1983; Evrard 1991; Bekki 1999;
Fujita et al. 1999).

The concordant cosmological model ($\Lambda$CDM) predicts the
formation and evolution of galaxy clusters through mergers of less
massive systems. Due to the large energies involved in cluster-cluster
collisions, a merging event can enhance the efficiency of the physical
mechanisms responsible for the evolution of the galaxy SFR.  The first
observational evidence for a correlation between cluster mergers and
SF activity came from the optical analysis of Coma, where an excess of
SF/PSF galaxies was observed in the collision region between the main
cluster and a group of galaxies (Caldwell et al. 1993). While in
several other non-relaxed clusters the collision between subclumps
seems to have increased the SF rate (SFR) of the galaxies
(e.g. Gavazzi et al. 2003), other analyses show the opposite trend
(e.g. Baldi et al. 2001). Therefore, the net role played by the
merging event on SF has still to be fully understood. In this picture,
a combined optical (Maurogordato et al.), X-ray (Sauvageot et al.) and
radio (Ferrari et al.) analysis of a sample of nearby clusters
($z\sim$0.09), covering all the main phases of the collision process,
has been undertaken. The main purpose of this project is to determine
the dynamical state of the observed clusters and understand if and how
the merging process affects the SF history of cluster members. In the
following, we will present the first analysis of the radio
observations (ATCA; Ferrari et al., in prep.) of the merging cluster
A3921 taking into account our previous optical results (Ferrari et
al. 2005, Paper I in the following).

\section{Dynamical state of Abell 3921}

A3921 ($z=0.094$) is a cluster with a perturbed morphology.  Recent
optical (Paper I) and X-ray analysis (XMM observations, Belsole et
al. 2005) have shown the presence of two dominant clumps of galaxies:
a main cluster centred on the BCG (A3921-A), and an NW sub-cluster
(A3921-B) hosting the second brightest cluster member. The comparison
of the optical and X-ray properties of A3921 suggests that A3921-B is
probably tangentially traversing the main cluster along the SW/NE
direction (Paper I; Belsole et al. 2005).

\section{Star formation properties}

The SF properties of the 104 confirmed cluster members in the central
field of A3921 (${\sim}1.8{\times}1.2 {\rm Mpc}^2$) have been
investigated by comparing: a) their spectral features (Paper I), and
b) their radio luminosities at 1.344 GHz derived from our new ATCA
observations (Ferrari et al., in prep). Based on the presence and
strength of the [OII]3727\AA~and Balmer lines, the galaxies of A3921
have been classified as (Paper I): a) {\bf k:} passive evolving
galaxies; b) {\bf k+a:} PSB(post-starburst)/PSF objects; c) emission
line galaxies, in turn divided in {\bf e(a):} dusty-SB and/or PSF
galaxies with some residual SF (Poggianti et al. 1999), {\bf e(b):}
SB's, and {\bf e(c):} classical spirals. It emerges that the
emission-line galaxies (11) share neither the kinematics nor the
projected distribution of the passive cluster members. Most of them
are spatially concentrated in the collision region of the two
subclusters, suggesting a possible connection between the detected
merger and the SF activity in A3921 (Paper I).

The radio luminosity is a dust-independent indicators of SFR, while
optical emission lines are not. We have therefore analysed the 1.344
GHz emission in the central field of A3921 in order to shed more light
on the SF properties of the confirmed cluster members. The SFRs of the
11 emission line galaxies, derived both from their [OII] emission
(Kennicutt 1998) and from their radio luminosities (Condon 1992), are
summarised in Table~\ref{tab:SFR}. At the mean cluster redshift and
with our cosmology (${\rm H}_0$=75~km/s/Mpc, $\Omega_m=0.3$ and
$\Omega_{\Lambda}=0.7$), our limit of radio detection at
3$\times$r.m.s. corresponds to a minimum detectable ${\rm
SFR}(M{\geq}0.1M_{\odot})$ of 3.69~$M_{\odot}{\rm yr}^{-1}$.  In
Table~\ref{tab:SFR} we also list the radio emitting cluster members
that do not show emission features in their spectra. They are all
k-type galaxies. No k+a galaxies have detectable radio emission at our
flux limit (3$\sigma$=0.147 mJy/beam, beam=$12''{\times}12''$).

\begin{table*}
\begin{center}   
\begin{tabular*}{1\textwidth}{@{\extracolsep{\fill}}lllccc}  
\hline
\hline
ID & ${\rm RA}_{\rm J2000}$ & ${\rm Dec}_{\rm J2000}$ & Spectral Type & ${\rm SFR}_{[OII]}$ & ${\rm SFR}_{\rm 1.344~GHz}$  \\
\hline
17 & 22 48 58.98 & -64 21 11.80	& e(a) & 0.70 & $<$3.69 \\
45 & 22 49 41.46 & -64 26 24.10 & e(c) & 0.10 & $<$3.69 \\
57 & 22 48 39.66 & -64 19 22.30	& e(c) & 0.61 & 8.78 \\
69 & 22 48 49.07 & -64 23 12.00 & e(b) & 14.09 & 48.79 \\
73 & 22 48 34.84 & -64 23 39.90	& e(c) & 0.64 & $<$3.69 \\
81 & 22 49 41.10 & -64 24 05.60 & e(a) & 0.12  & $<$3.69 \\ 
82 & 22 49 38.41 & -64 23 23.80 & e(a) & 0.40 & $<$3.69 \\
100 & 22 49 41.58 & -64 19 59.00 & e(b) & 1.74 & $<$3.69 \\
169 & 22 50 01.83 & -64 22 21.80 & e(a) & 0.31 & $<$3.69 \\
181 & 22 49 26.08 & -64 23 21.40 & e(c) & 0.24 & $<$3.69 \\
226 & 22 49 11.96 & -64 16 03.90 & e(a) & 0.43 & $<$3.69 \\
\end{tabular*}
\begin{tabular*}{1\textwidth}{@{\extracolsep{\fill}}lllccccc}
\hline
\hline
ID & ${\rm RA}_{\rm J2000}$ & ${\rm Dec}_{\rm J2000}$ & Spectral Type & log${\rm L}_{\rm 1.344~GHz}$ & & & \\
\hline
39 & 22 50 06.48 & -64 24 41.90	& k & 24.07 & & & \\
41 & 22 49 58.18 & -64 25 48.10	& k & 22.28 & & & \\
52 & 22 49 04.78 & -64 20 35.60	& k & 23.89 & & & \\
71 & 22 48 43.36 & -64 23 53.10	& k & 21.28 & & & \\
94 & 22 49 58.14 & -64 20 05.80	& k & 22.10 & & & \\
\hline
\end{tabular*}
\caption{{\bf Top:} SFR ($M_{\odot}{\rm yr}^{-1}$ for
$M\geq0.1M_{\odot}$) of the emission line cluster members measured
through their 1.344~GHz and [OII] luminosities. ID's correspond to
Tables A in Paper I. {\bf Bottom:} 1.344 GHz luminosities (${\rm
W}{\rm Hz}^{-1}$) and spectral types of the radio emitting galaxies in
A3921 central fields, in which no emission lines have been detected. }
\label{tab:SFR}
\end{center}
\end{table*}

\section{Discussion and conclusions}

{\bf - Emission line galaxies:} only two have been detected on the
1.344 GHz map. One (\#69) is most likely a Seyfert galaxy (radio and
X-ray emission, strong [OII] emission); the other (\#57) is a
classical spiral (typical: spectral properties, SFR, morphology and
colour), whose SFR in optical (Table~\ref{tab:SFR}) is underestimated,
probably due to dust obscuration. The remaining (9) emission line
galaxies are located in the centre of the cluster, and they have a
${\rm SFR}_{\rm [OII]}$ much lower than the minimum SFR detectable
with our radio observations. Since none of them has actually been
detected at radio wavelengths, we confirm their low SFR ($<$3.7
$M_{\odot}{\rm yr}^{-1}$) without bias due to dust
obscuration. Therefore they could be gas-poor galaxies with very weak
SF. This would confirm our hypothesis of Paper I, i.e. not all the SF
galaxies in the collision region of A3921 are infalling, gas-rich
field galaxies. They could be really located at the cluster centre
(and not seen in projection), and their ongoing SF could have been
refuelled by the cluster merger.
\smallskip\\ {\bf - k+a galaxies:} non-detection at radio wavelengths
confirms that they are PSF galaxies.
\smallskip\\ {\bf - Radio emitting galaxies without emission lines:}
their radio luminosities and spectral types suggest that their radio
emission is probably due to a central AGN, rather than to
SF\footnote{The faintest source (\#71) could be either a weak AGN or a
SF galaxy (obscured by dust in optical).}. Among them, one (\#41) is
associated with the BCG; one, with a distorted radio morphology, to
the second BCG (\#52); one is located at the cluster centre (\#39),
and it shows a head-tail morphology possibly associated to the
relative motion of the radio source with respect to the dense ICM.

We therefore conclude that SF has been triggered in a fraction of
cluster galaxies located in the collision region of A3921.



\end{document}